\def\simgr{\,\hbox{\hbox{$ > $}\kern -0.8em \lower 1.0ex\hbox{$\sim$}}\,}
\def\simle{\,\hbox{\hbox{$ < $}\kern -0.8em \lower 1.0ex\hbox{$\sim$}}\,}
\documentclass{aa}  
\usepackage{graphicx}
\usepackage{natbib}
\usepackage{txfonts}
\def\simgr{\,\hbox{\hbox{$ > $}\kern -0.8em \lower 1.0ex\hbox{$\sim$}}\,}
\begin{document}
   \title{Lithium depleted stars in the young $\sigma$ Ori cluster\thanks{Based
   on data
   collected at the ESO Very Large Telescope, Paranal Observatory, Chile [program
   074.D-0136(A)]}}

   \subtitle{}

  \author{G. G. Sacco 
          \inst{1}\fnmsep\inst{2}
         \and S. Randich\inst{3}
	  \and
         E. Franciosini\inst{1}
	   \and
	  R. Pallavicini\inst{1}
	  \and
	  F. Palla\inst{3}}

   \offprints{G. G. Sacco, e-mail: sacco@astropa.unipa.it}

   \institute{INAF-Osservatorio Astronomico di Palermo, Piazza del Parlamento,
   1, 90134 Palermo, Italy\\
              \and Dipartimento di Scienze Fisiche ed Astronomiche, Sezione
	 Astronomia, Piazza del Parlamento, 1, 90134 Palermo, Italy\\
           \and INAF-Osservatorio Astrofisico di Arcetri, Largo E. Fermi, 5, 50125
	   Firenze, Italy }

   \date{Received 22 september 2006; accepted 27 november 2006}

 
  \abstract
   {Knowledge of the age distribution of stars in young clusters
    and associations
   is crucial to constrain models of star formation.
   HR diagrams of different young clusters and associations 
   suggest the presence of age spreads, but 
   the influence of errors on the derived ages is still largely 
   debated. 
   Determination of lithium abundances 
   in low-mass 
   stars represents an alternative and robust
    way to
   infer stellar ages.}
   {We measured lithium in a sample of low mass members
    of the young (4-5 Myr) $\sigma$ Ori  
   cluster with the main goal of investigating its star formation history.}
   {Using the FLAMES multi-object spectrograph on VLT/UT2, 
   we obtained spectra of 98 candidate cluster members. The
   spectra were used to determine radial velocities, to infer the presence of
   H$\alpha$ emission, and to measure the strength of the Li~{\sc i}
   670.8~nm absorption line.}
   {Using radial velocities, H$\alpha$ and Li, together with information
   on X-ray emission, we identified 
   59 high probability cluster members. Three of them show severe Li depletion. 
    The nuclear ages inferred for these highly depleted stars
   exceed 10-15~Myr; 
   for two of them these values are in good agreement with the isochronal
   age, while for the third star the nuclear age exceeds 
   the isochronal one.} 
   {}

   \keywords{stars: pre-main sequence-- 
   Open clusters and associations:$\sigma$ Ori--stars:abundances                             }

 \maketitle
%

\section{Introduction}

The time scale of  stellar birth within molecular clouds 
is one of the main open issues in star formation theory.
Two scenarios have been proposed to explain the
observational results: Rapid Star Formation (RSF) or  
Slow Accelerating Star Formation  (SASF) 
\citep{Elmegreen2000ApJ, Palla2002ApJ}. 
HR diagrams, which are the main tool 
to determine Pre-Main Sequence (PMS) stellar 
ages, show a spread of $\simgr$10$^7$ years 
in individual clusters and associations,
supporting the SASF model, 
but the effect of various sources of errors, 
both observational and theoretical, 
is largely debated \citep{Hartmann2001AJ, Burningham2005MNRAS}. 
Lithium (Li) abundances can be used as an independent and robust method 
to determine ages of young
objects \citep{Martin1998MNRAS,Martin1998AJ}, since low-mass stars in 
the range 0.5--0.08 $M_{\odot}$ deplete their initial
Li content during the PMS phase 
\citep{Bodenheimer1965ApJ, Baraffe1998A&A}. The timescale
of Li depletion depends on mass,
with higher mass stars 
(0.5-0.3 M$_{\odot}$) starting to burn it after 5-10 Myr and 
lower mass stars (M$<$0.2 M$_{\odot}$) after 20-30 Myr.
\cite{Palla2005ApJL} 
employed the Li age-dating method among low-mass members 
of the Orion Nebula Cluster
and found four Li-depleted stars with nuclear
ages $\sim 10$~Myr.
 
The $\sigma$ Ori cluster was discovered by ROSAT \citep{Walter1997MmSAI} 
around the O9.5 V binary star $\sigma$ Ori AB (distance 350$^{+166}_{-85}$ pc,
\citealt{Perryman1997A&A}).
Its low-mass stellar population
has been intensively studied by photometry and low-resolution 
spectroscopy in the optical and near-infrared
\citep{Zapatero2002A&A, Barrado2003A&A, 
Sherry2004AJ, Scholz2004A&A,Oliveira2004MNRAS},  while the 
X-ray properties have been investigated most recently by 
\cite{Franciosini2006AA} using XMM-Newton.

\citet{Oliveira2002A&A} determined an isochronal median age of
4.2$^{+2.7}_{-1.5}$ Myr. Subsequently, \citet{Oliveira2004MNRAS} presented 
evidence  for  a large age spread ($\sim$30 Myr) in the I/I-J Color-Magnitude  
Diagram (CMD), and suggested that part of it could be due 
to photometric variability of the PMS stars. 
\citet{Kenyon2005MNRAS} have found {\it bona fide} cluster members
with small Li line equivalent widths (EW), possibly indicating
a certain amount of depletion.
Finally, \citet{Jeffries2006MNRAS} have discovered two kinematically separate 
populations of young PMS stars, one concentrated around $\sigma$ Ori AB 
sharing a common radial velocity with this star 
({v$_1=31.0\pm 0.5$ km/s)},
and the second one, more dispersed in the sky, 
with a radial velocity similar to that of the Orion OB1a and 1b associations
({v$_2=23.8\pm 1.1$ km/s)}.

In this Letter, we report on the discovery of three high-probability members 
of the main $\sigma$ Ori cluster with Li abundances a factor of about 1000
below the interstellar value. This
result was obtained as part of a VLT/FLAMES 
survey to study membership, Li abundances and accretion diagnostics of a large 
sample of K and M stars around $\sigma$ Ori AB (Sacco et al., in preparation).

\section{Observations and Results}   
     
\subsection{Target selection and observations}

We have selected 98 cluster candidates from previous studies. 
In order to secure
a high fraction of cluster members, 
we have granted higher priority to stars detected in X-rays
by \citet{Franciosini2006AA} and with isochronal ages $\simle$10~Myr 
from optical and infrared CMDs.
All stars of the sample have an infrared counterpart in the 2MASS
catalog \citep{Skrutskie2006AJ}, while for 79 of them optical photometry
is available in the literature 
\citep{Wolk1996PhDT, Zapatero2002A&A,Barrado2003A&A, Burningham2005MNRAS,
Kenyon2005MNRAS}.

The sample stars were observed using FLAMES on VLT/UT2
\citep{Pasquini2002Messanger}. FLAMES was operated in MEDUSA mode 
with the high resolution ($\lambda/\Delta\lambda$=17000) 
HR15N grating covering the spectral range 647-679~nm. The
Field of View (FoV) was centered at
RA=05$^{h}$38$^{m}$48$^s$.9 and Dec=-02$^{d}$34$^{m}$22$^s$ and is almost 
coincident with the FoV of XMM-Newton \citep{Franciosini2006AA}. 
The observations were divided in six runs between October 
and December 2004, for a total exposure time on target of 4.3 hours. 
Data reduction was performed using the GIRAFFE girBLDRS pipeline 1.12, 
following the standard steps \citep{Blecha2004Man}. Sky subtraction was 
performed separately using the average of six-seven sky spectra.

\subsection{Membership}

\begin{figure}[!h]
\centerline{ \hbox {
\includegraphics[height=6 cm, angle=0]{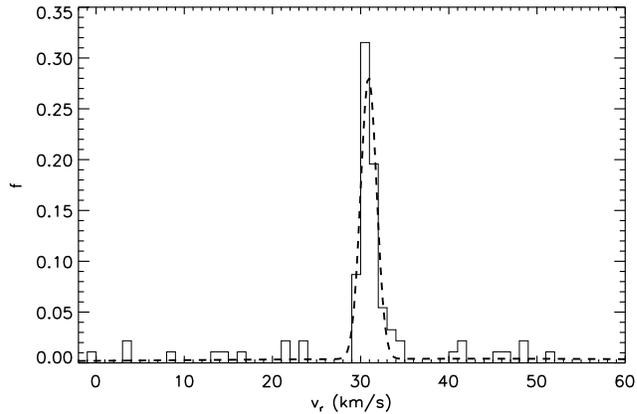} }}
\caption{Observed RV distribution of all the stars of the sample, with 
overplotted (dashed line) the curve resulting from the fitting with the sum 
of two gaussians.}
\label{fig:RV_dist}
\end {figure}

We have measured radial Velocities (RVs) using the IRAF\footnote{IRAF is 
distributed by the National Optical Astronomy Observatory, which is operated by
the Association of Universities for research in Astronomy, Inc.,under contract to
the National Science Foundation.} task FXCOR 
by Fourier cross-correlation with two template spectra
chosen among sample stars with no accretion signatures.
For 86 stars we derived a median RV as the weighted average of the 
six values measured for the different runs, while the remaining 12 stars turned out to be candidate
binaries and will be discussed in the forthcoming paper.    
Cluster RV was determined by
fitting the observed RV distribution, shown in Figure \ref{fig:RV_dist},
 with the weighted sum of 2 gaussian distributions, one
for the cluster and the other for the field. The best fit yields a value of
v$_C$=30.91$\pm$0.90 km/s for the cluster
and v$_F$=43$\pm$ 36 km/s for the field. 
Our cluster velocity is in excellent agreement
with that (v$_C$=31.0$\pm$0.5 km/s) determined by \citet{Jeffries2006MNRAS}.

We considered as cluster members stars with RVs within  
3~$\sigma$ from the average, yielding
61 members and 25 field stars. The statistical contamination of the cluster
member sample, estimated by integrating the field star distribution between 
v$_{C}-3\sigma_{C}$ and v$_{C}+3\sigma_{C}$, is $\sim$2 stars.
The second requirement on membership is the presence of H$\alpha$ in emission.
Among the 25 RV non-members, 24 are field 
stars also according to H$\alpha$, 
while one star, which appears young based on
H$\alpha$, has a radial velocity (v=23.73$\pm$0.45 km/s) similar to that 
of the second population discovered by \citet{Jeffries2006MNRAS}. 
Note that all the non-members have Li~I 670.8 nm pseudo Equivalent Width
(pEW) smaller than 200 m\AA.
Finally,
among the 61 RV cluster members, 56 are PMS stars also
from H$\alpha$ emission and from the strength of the Li line; 
two stars with 
H$\alpha$ in absorption and no Li are most likely field stars. 

\begin{figure}[!h]
\centerline{ \hbox {
\includegraphics[width=9 cm, angle=0]{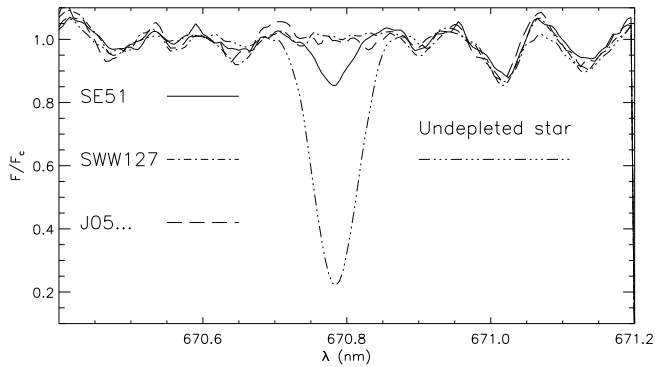} }}
\caption{Spectra of the three Li-depleted stars listed in Table \ref{tab:info};
namely,
SWW127 (dotted line), J053914.5-022834 (dashed line), and SE51 (continuous
line). For comparison 
the spectrum of an undepleted star in same temperature range of 
those reported in Table \ref{tab:lidep} is also shown (dotted-dashed line)}
\label {fig:spectra}
\end {figure}

\begin{table}
\begin{center}
\caption{\label{tab:info} Membership indicators of Li depleted stars.}
\begin{tabular}{ccccc}
\hline
\hline
Object &v$_r$ & H$\alpha$ EW  & Log L$_X$ & PMS \\
& (km/s) & (\AA) & (erg/s) &indicator \\
\hline
SE51$^a$ & 29.5$\pm$0.5  &$-$1.1   & 29.22  &variable  \\
SWW127$^b$& 33.2$\pm$0.5   &$-$2.6 & 29.51 &  -\\
J053914.5-022834$^c$ & 29.2$\pm$0.5  &$-$4.6 & 30.17&NIR ex. \\
\hline
\multicolumn{5}{l}{$^a$~\citet{Scholz2004A&A};$^b$~\citet{Sherry2004AJ};} \\
\multicolumn{5}{l}{$^c$~\citet{Zapatero2002A&A}.}\\
\end{tabular}
\end {center}
\end{table}

The remaining 3 RV cluster members, listed in Table \ref{tab:info}, have 
H$\alpha$ in emission and  Li pEW less than 200 m\AA. 
As shown in Figure \ref{fig:spectra}, 
the Li line is clearly 
identified (EW=150 m\AA) in the spectrum of 
SE51, but not in those of SWW127 and 
J053914.5-022834. These stars have an X-ray counterpart and
result to be PMS stars in CMDs. Moreover, J053914.5-022834 
shows excess in K-L \citep{Oliveira2004MNRAS} and SE51 is characterized 
by photometric 
variability \citep{Scholz2004A&A}.
Considering that only two out of 22 field stars in our sample
have an X-ray counterpart (2 field stars are out of the XMM FoV) and that the probability of 
having a field star with RV between v$_C-3\sigma_C$ and v$_C+3\sigma_C$ 
is 0.056, the probability of finding, among the RV cluster members,
3 field stars with an X-ray counterpart is less than 5$\times$10$^{-3}$.  We conclude that, although physical membership must be confirmed by, e.g., proper motion studies, we regard as unlikely that all of them are non members.

\begin{figure}[!h]
\centerline{ \hbox {
\includegraphics[width=9 cm, angle=0]{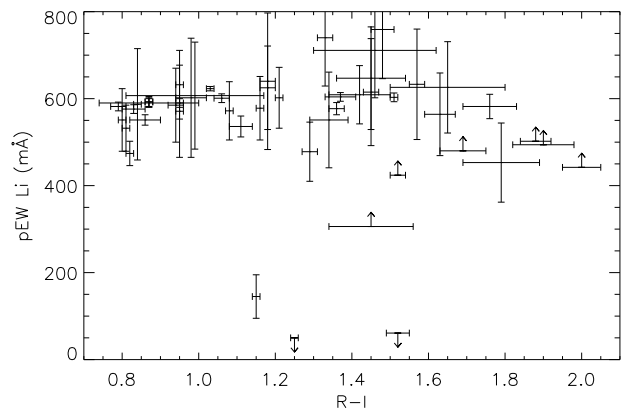} }}
\caption{pEWs of the Li line corrected for veiling as a function of R--I
color for cluster members with
available optical photometry. Upward and downward arrows denote lower and upper
limits.}
\label {fig:pew}
\end {figure}

\subsection{Lithium depleted stars}  

pEWs of the Li~line
were measured using the IRAF task SPLOT, by direct integration
of the line profile over an interval of 0.2~nm. 
Measured pEWs may be affected by spectral veiling.
We have estimated $r$, the ratio of the excess emission to the photospheric
continuum, for all stars of the sample
from the measurement of the EWs of three absorption lines (V~I~662.5~nm,
Ni~I~664.3 nm, Fe~I~666.3~nm) in their spectra 
and in those of 11 unveiled comparison stars
with similar temperatures (see \cite{Palla2005ApJL}).
In those cases where we were not able to measure at 
least 2 lines, because of low S/N or high veiling, we have considered 
$r$ indeterminate and the measured Li pEWs are lower limits
to the true values. 
Derived $r$ values range between 0 and 1.4, with errors of about
0.1-0.2. Figure \ref{fig:pew} shows pEWs of the Li line, corrected for veiling,
for stars with available optical photometry. The median pEW is 590 m\AA, 
with a typical dispersion of $\sim$100 m\AA.
The scatter in Li pEWs could be due to measurement errors, but we 
cannot exclude the presence of some partially depleted stars. 
The three stars reported in Table \ref{tab:info} are not veiled ($r$=0.0-0.2)
and their Li pEWs are less than 200 m\AA, indicating a large amount of Li
depletion.

\begin{table*}
\begin{center}
\caption{\label{tab:lidep} Li depleted stars and their properties.}
\begin{tabular}{ccccccccc}
\hline
\hline
 Object & T$_{\rm eff}$ (K)  & Log(L/L$_{\odot}$)  & Li EW  & A(Li)& M$_{\rm HRD}$ & t$_{\rm HRD}$ & M$_{\rm Li}$ & t$_{\rm Li}$\\
 & (R,I/I,J)  & (R,I/I,J) & (m\AA) &  & (M$_{\odot}$) & (Myr) & (M$_{\odot}$) & (Myr)\\
\hline
SE51 & 3750$\pm$20/3750$\pm$60  & -0.91$\pm$0.01/-0.91$\pm$0.02& 150$\pm$50 &$-$0.5$\pm$0.5&   
0.52--0.53 & 12--13 & $0.49 \pm 0.04$ & $10.5 \pm 0.6$\\
SWW127 &  3645$\pm$15/3720$\pm$60 &-0.92$\pm$0.01/-0.94$\pm$0.02& $<$90 &$<$-1.0 & 0.42 -- 0.49 & 8.4-14.2 &
$\geq 0.44 $ &  $\geq 10$\\
J053914.5-022834 & 
3345$\pm$70 & -1.14$\pm$0.06 & $<$100 &$<$-0.7 & $0.23 \pm 0.05$ &
$4.5 \pm 1.5$ & $\geq 0.4$ & $\geq 15 $\\
\hline
\end{tabular}
\end {center}
\end{table*}

We have determined Li abundances for these three stars using Curves of Growths 
of \citet{Palla2006A&A}, 
based on \citet{Pavlenko2001AstRep} models and spectral code.
Surface gravity was fixed at $\log$~g=4.0~dex, the expected value for a
4 Myr old late type star by \cite{Baraffe1998A&A} models.
The effective temperature of the depleted stars is given
in Table~\ref{tab:lidep}: in the case of J053914.5-022834 
it has been derived from the spectral
type measured by \citet{Zapatero2002A&A}, 
while T$_{\rm eff}$ of SE51 and SWW127 are
estimated from the color indexes R-I and I-J, using the models of
\citet{Baraffe1998A&A}.
The resulting Li abundances are listed 
in Table~\ref{tab:lidep} and are a factor
of $\simgr$~1000 below the interstellar value (A(Li)=3.3).
We stress that even relatively large errors in T$_{\rm eff}$ 
and/or log~g would not greatly affect Li abundances (or depletion factors),
given the extremely small pEWs (or upper limits to
pEWs). Similarly, the low abundances cannot be explained
  by veiling, since it would imply a value of $r\gg 1$, or by
  other sources of error, such as uncertainty in the pseudo-continuum.
  Contrary to the suggestion of \citet{Kenyon2005MNRAS}, we believe that the
  measured low pEWs represent real Li depletion, since
  Pavlenko (2001) models provide a correct treatment of the Li line 
  formation in cool objects and pEWs below $\sim$170~m\AA~imply that only TiO 
  contributes to the 670.8~feature.

\section{Discussion and Conclusion}

\begin{figure}[!h]
\centerline{ \hbox {
\includegraphics[width=8.0 cm, angle=0]{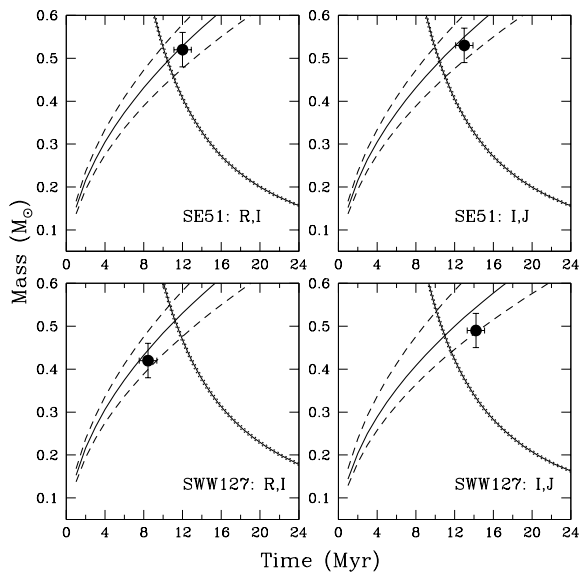} }}
\caption{Mass vs. age plot for the Li depleted stars SE51 and SWW127.
T$_{\rm eff}$ and bolometric correction have been derived
using the color indices R-I (left-hand panels) and I-J (right-hand panels)
and the models of \citet{Baraffe1998A&A}.
Continuous curves represent the locus of constant luminosity (positive slope) 
and constant Li abundance (negative slope), while dashed curves indicate the
uncertainty range. The circles with the error bars give the masses and ages
computed from models of \citet{Palla1999ApJ}.}
\label {fig:bild2}
\end {figure}

The three Li depleted, high probability members of the $\sigma$ Ori cluster represent an
example of post T Tauri stars (following the definition
of \citet{Martin1997A&A}),  and allow us to investigate
its star formation history from a novel
point of view.
Isochronal and nuclear ages (t$_{\rm HRD}$) and masses (M$_{\rm HRD}$) 
for the three Li-depleted stars 
are listed Table~\ref{tab:lidep}. These values 
have been derived from the models of \citet{Palla1999ApJ} using the 
T$_{\rm eff}$ and luminosity range given in the table. 
In the last two columns we list the nuclear mass (M$_{\rm Li}$) and 
age (t$_{\rm Li}$) estimated following ~\cite{Palla2005ApJL}, based on 
\cite{Bildsten1997ApJ}. The latter provides analytical prescriptions
to derive the age vs. stellar luminosity and the age vs. Li 
abundance relations for a fully convective star undergoing 
gravitational contraction at a nearly constant T$_{\rm eff}$ and assuming fast 
and complete mixing. In Figure \ref{fig:bild2} we display 
the mass vs. age relations at fixed luminosity (positive slope) and 
Li abundance (negative slope) 
predicted by the models. The intersection of the two curves 
gives the mass and nuclear age of the star. 

We find that
the age and mass of SE51 and SWW127 derived from Li
are in very good agreement with the isochronal values obtained
using both R--I and I--J colors. 
On the contrary, the measured amount of Li depletion for 
J053914.5-022834 is too large for the isochronal age and low mass from
evolutionary tracks.

Considering that most of the remaining stars of our sample have pEWs consistent
with the interstellar value, our results support the view that the bulk of 
the $\sigma$~Ori population has an age $\simle$~4--6 Myr.
However, three low-mass, high probability members 
are definitively older than $\sim$10 Myr.
Therefore, we propose that $\sigma$ Ori has been forming stars
on a time scale of $>$10-15 Myr,
as found in the case of the Orion Nebula Cluster.

Finally, we note that our sample has been selected requiring an 
isochronal age from CMDs less than $\sim$10 Myr.
This bias has precluded the identification of 
additional Li-depleted, old stars that might exist in the $\sigma$ Ori cluster.
Further observations at low spectral resolution to derive stellar parameters, 
as well as at high resolution to measure lithium abundances on a larger sample 
of stars can help to fully characterize the star formation history of the 
cluster.

\bibliographystyle{aa}
\bibliography{biblio}

\end{document}